# Mid-board miniature dual channel optical transmitter MTx and transceiver MTRx

**Xiandong Zhao[a], Chonghan Liu[a], Datao Gong[a*], Jinghong Chen[d], Di Guo[a], Huiqin He[f], Suen Hou[c], Guangming Huang[b], Xiaoting Li[b], Tiankun Liu[a], Xiangming Sun[b], Ping-Kun Teng[c], Le Xiao[b,a], Annie C. Xiang[a], Jingbo Ye[a]**

[a] *Department of Physics, Southern Methodist University, Dallas, TX 75275, USA*

[b] *Department of Physics, Central China Normal University, Wuhan, Hubei 430079, P.R. China*

[c] *Institute of Physics, Academia Sinica, Nangang 11529, Taipei, Taiwan,*

[d] *Department of Electrical and Computer Engineering, University of Houston, Houston, TX 77004, USA*

[e] *State Key Laboratory of Particle Detection and Electronics, University of Science and Technology of China, Hefei Anhui 230026, P.R.China*

[f] *Shenzhen Polytechnic, Shenzhen 518055, P.R.China*

[*] *dgong@mail.smu.edu*)

ABSTRACT: We report the development of a mid-board, TOSA and ROSA based miniature dual channel optical transmitter (MTx) and a transceiver (MTRx). The design transmission data rate is 5.12 Gbps per channel and receiving data rate 4.8 Gbps. MTx and MTRx are only 6 mm tall and are electrically and optically pluggable. Although the fiber TOSA/ROSA coupling is through a custom latch, the fiber uses the standard LC ferrule, flange and spring. Light coupling is ensured by the TOSA and ROSA with the LC coupling mechanism. With the dual channel serializer LOCx2 sits under MTx, one achieves high data transmission with a small PCB footprint, and enjoys the reliability of the hermetically packaged TOSA. MTx and MTRx are designed for detector front-end readout of the ATLAS Liquid Argon Calorimeter (LAr) trigger upgrade.

**Contents**

**1. Introduction**

We develop the mid-board miniature dual channel optical transmitter (MTx) and transceiver (MTRx) for detector front-end readout of the ATLAS Liquid Argon Calorimeter trigger upgrade [1] where existing mechanical constraint requires these optical modules to be at most 6 mm tall that is almost the diameter of a TOSA or ROSA. Many multi-giga-bit-per-second readout channels on a motherboard require mid-board optical transceivers to pick the fast signal where the serializer or ser-des is and to route the signals optically to/from the front-panel. With the dual channel serializer ASIC LOCx2 [2] that fits under MTx, the footprint of the serial data transmission on a PCB is minimized. TOSA and ROSA, that have the standard LC ferrule, flange and spring assembly, are used with fibers to ensure light coupling efficiency while minimize custom made parts to control cost. Hermetically packaged TOSA and ROSA offer operational reliability. MTx and MTRx are electrically and optically pluggable, making production testing, system installation and maintenance easy. MTx and MTRx are complementary to the panel-mount VTTx and VTRx optical modules that are developed by the Versatile Link project [3].

**2. The design of MTx and MTRx**

The key design points in MTx and MTRx are a custom latch that couples the fiber to TOSA/ROSA, and attach the TOSA/ROSA to the PCB in which there is a cutout to make MTx and MTRx only as tall as the diameter of TOSA/ROSA. The electric coupling is through the connector (Part number LSHM-120-02.5-L-DV-A-N-TR) from Samtec that has a 5 mm stacking height. To operate in the detector front-end of ATLAS LAr, the VCSEL driver is LOCld [4] and the p-i-n diode TIA is GBTIA [5]. One can also use COTS drivers and TIAs should the operation environment allow.

Shown in Figure 1 is the design of the custom latch with the way it couples the TOSAs and the fibers together. The latch has two parts. The first one holds the TOSA/ROSA and anchors them to the PCB with a screw through the hole in the center beam. That screw also doubles as the anchor of MTx or MTRx to the motherboard. The second part, when engaged, pushes the LC ferrules into the TOSA/ROSA through the springs. The latch is injection molded with the same material that makes the body of the TOSA/ROSA. Shown in Figure 2 is the fiber with the LC ferrule, flange and spring, all standard and can be ordered from manufacturers that assemble standard LC connectors. The only difference of this assembly from the standard LC connector is not to add the bulky LC connector housing after the ferrule and flange are attached to the fiber. As MTx and MTRx are meant for mid-board connections, frequent plug and unplug are not foreseen.

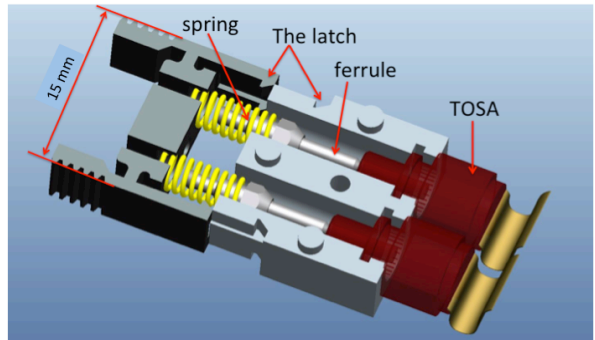

**Figure 1:** The design of the latch and the coupling of fiber to the TOSA/ROSA.

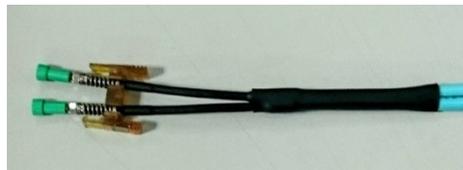

**Figure 2:** The fiber with the second part of the latch attached.

Shown in Figure 3 is the way MTx or MTRx is mounted on a motherboard. The height of MTx or MTRx from the motherboard to the top surface of the optical module is only 6 mm. There is no component on the top of the module PCB.

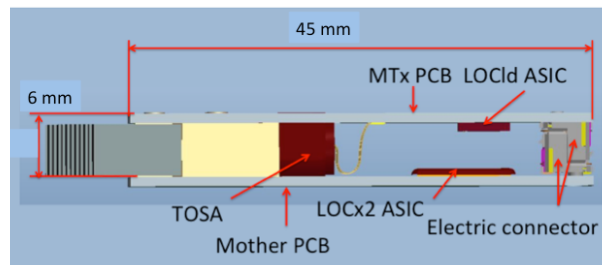

**Figure 3:** An MTx mounted on a motherboard with the dual channel serializer ASIC LOCx2 under it.

### 3. MTx and MTRx test results and status

Shown in Figure 4 are a measured optical eye diagram of MTx, and an electrical eye diagram of MTRx (the receiving channel). Both eyes are wide open at 5 Gbps.

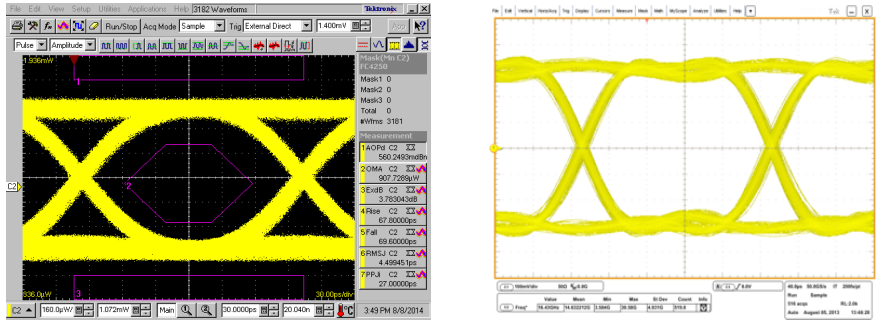

**Figure 4:** Left: the optical eye from MTx and the transmitting channel of MTRx. Right: the electrical eye from the receiving channel of MTRx.

The power consumption of the MTx with LOCld is about 200 mW per module and MTRx 320, making both suitable for operations with cooling with only ambient air. Long-term reliability of these modules relies on the PCB assembly that includes manual soldering of the TOSA/ROSA, and the components in the module. Accelerated life test is being performed [6].

Shown in Figure 5 is a photo of MTx prototype. All the modules meet the design specifications to operation at 5 Gbps and in the environment of ATLAS LAr detector front-end. Production will start after the life test and the production readiness review in ATLAS.

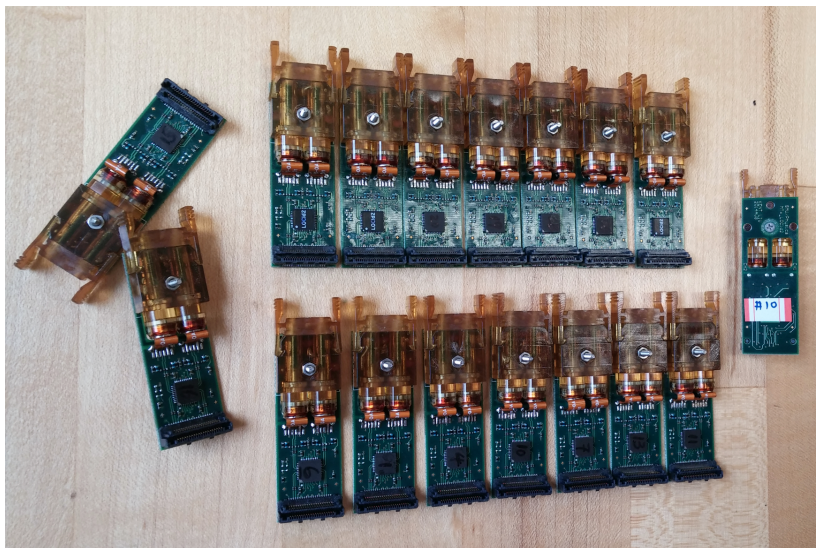

**Figure 5:** Photo of MTx prototypes.

## 4. Conclusion

We have developed a mid-board miniature dual channel optical transmitter MTx and a transceiver MTRx, using standard LC TOSA/ROSA coupling LC ferrule fibers. The coupling is achieved by a custom latch that also anchors the TOSA/ROSA to the module PCB and the mother-PCB. MTx and MTRx operate at about 5 Gbps each channel. These mid-board optical modules are designed for ATLAS LAr trigger upgrade detector front-end readout. They can also be used in other systems where mid-board optical coupling is needed. Both MTx and MTRx will soon be ready for production.

**Acknowledgments**

This work is supported by the US-ATLAS R&D program for the upgrade of the LHC, the US Department of Energy under the Grant No. DE-FG02-04ER1299, and the CDRD grant from DOE. The idea of coupling a ferrule directly to a TOSA came initially from a presentation made by Csaba Soos (CERN) in a Versatile Link collaboration meeting. The coupling used in MTx and MTRx is developed from that idea. The authors would like to express their deepest appreciation for Csabe Soos and colleagues in the Versatile Link collaboration for this inspiration. We also would like to thank Prof. Ming Qi of Nanjing University for the help on injection molding.